# USABILITY ENGINEERING OF GAMES: A COMPARATIVE ANALYSIS OF MEASURING EXCITEMENT USING SENSORS, DIRECT OBSERVATIONS AND SELF-REPORTED DATA

Arwa Alamoudi[1], Noura Alomar[1], Rawan Alabdulrahman[1],
Sarah Alkoblan[1] and Wea'am Alrashed[1]

[1]Department of Software Engineering, King Saud University, Riyadh, Saudi Arabia

`{aalamoudi, nnalomar, ralabdulrahman, salkoblan, wealrashed}@ksu.edu.sa`

*ABSTRACT*

*Usability engineering and usability testing are concepts that continue to evolve. Interesting research studies and new ideas come up every now and then. This paper tests the hypothesis of using an EDA-based physiological measurements as a usability testing tool by considering three measures; which are observers' opinions, self-reported data and EDA-based physiological sensor data. These data were analyzed comparatively and statistically. It concludes by discussing the findings that has been obtained from those subjective and objective measures, which partially supports the hypothesis.*

## 1. INTRODUCTION

Usability testing has emerged as an essential phase in software development and it examines the property of systems being usable. This paper concentrates on taking the advantage of the EDA-based physiological measurements, which has the ability to gather data from physiological responses of the body. Consequently, the main idea behind this study is to test the hypothesis of using the EDA-based sensors as a usability testing tool. An experiment has been conducted to see if the findings can support this assumption. In this experiment, the participant wears the sensor bracelet while playing a game in order to perform physical activities and thus increasing the arousal level of the player. Three usability measures were considered in this experiment; external observation; external self-reporting; and internal reactions measured by the sensor.

This paper is organized into six sections. Introduction is shown in section I. In section II, the literature review is presented. Then, the experiment's design is discussed. The result obtained in this research study is provided in section IV. After that, the discussion of the result is shown in section V. Lastly, section VI concludes the paper.

## 2. RELATED WORK

### 2.1 Usability Testing

Usability testing is the process of measuring the quality of the system being tested by gaining the users' feedback while they are interacting with the system and using its services [1]. The purpose of this testing is to understand and uncover the defects that might be encountered by the users after releasing the system and to obtain the users' satisfaction. This can be done by preparing the test plan, setting up the environment properly and using either traditional usability testing methods. Some examples of the methods are observing the participants and asking them to fill out a survey on site or by using software or hardware tools that are developed specifically for this purpose and can help in performing the usability tests. Moreover, some of these tools automate the external observation process while others deal with the physiological issues that occur inside the bodies of the participants. Choosing the usability testing techniques and tools depends on the type of the system being tested and the type of information that need to be gathered. For instance, nowadays mobile applications are used frequently without being restricted to any place such as offices or labs. In this case, usability testing practitioners cannot use the observation method because they need to test the device or application and use a data collection method that can work properly in a non-controlled environment. Consequently, the different states that the user might have when using the mobile device such as walking on the street and driving a car need to be considered [11]. Researchers have proposed ways to address the mobility nature, [10] evaluated six situations by conducting two experiments on people while they are using their smart phones. Hence, by considering the EDA-based sensors as a usability testing tool, the goal of this research is to know if the EDA sensors are as effective as external observation. As a result, the following sections examined the literature from three angles which are the external observation, affective computing aspects and internal body changes.

Many studies have been concerned with comparing the usability assessment tools and measuring the accuracy of them. For example, [1] presents a usability testing study that was performed to benchmark a developed system called Ultimate Reliable and Native Usability System (URANUS) against other software tools which are Morae, Tobii, Userfeel and Loop1. And by comparing their features, this study reports that Morae achieves the highest score. On the other hand, the authors state that Tobii Studio has the advantage of using the eye-tracking feature which help in capturing the eye movements of the participant [1]. Another user experience study has been conducted to compare the interface design of two portable ultrasound scanners [7]. In this study, cameras and eye-tracking device were used to observe the sonographers' eyes and upper-body kinematic while they were using the scanners. They were able to conclude that there was a difference between the left hand movements of the two participants. In addition, the eye-tracking analysis shows that there were no differences between the participants' eyes movements. Clearly, the findings were based on the analysis of the eye-tracker data and the cameras' photos without any supporting information from the participants and this means that the eye-tracker has been used as a usability testing tool. In another study [8] they combined data from eye-tracker with subjective satisfaction measures. This paper discusses this point by using the eye tracker to measure the fixation frequency and fixation duration of the participants while they were performing tasks on the Web. After that, a comparison between the data generated from the eye tracker and the self-reported data from the collected surveys confirms that there was a correlation between these objective measures and subjective measures [13].

In usability testing sessions, the existence of facilitators to observe the participants and collect information about their performance manually by filling out pre-designed forms has proven to be useful to the usability testing practitioners. For instance, the observation technique has been used in [9] to evaluate the usability of educational game designed for children. In this experiment, the children were observed by a facilitator to know their body language, facial expressions and any other feedback that they show while they were playing the game. Therefore, the results were driven based on the information collected through the observation and game experience questionnaires filled out by the participants.

Using objective measures in user experience testing (UET) involves studying the changes in the physiological factors of the participants' bodies. Evidence suggested that Biofeedback data analysis has comparable results when compared to the subjective analysis of paper surveys. The objective of that experiment carried out in this research was to see if the effects of the different color combinations of texts and backgrounds of a website and the mental stress happened in each case can be reflected by the physiological analysis [12]. Moreover, [13] suggested using a triangular approach in UET which was based on using traditional usability testing methods and the self-reported data combined with physiological and neurological measurements.

## 2.2 Affective Computing

Examining the usability of the interface designs through measuring the emotions that users show to computers is a young field of research. This ability of extracting the emotions of users in certain situations comes under the field of affective computing. Affective computing can help human-computer interaction specialists, particularly in the usability and intuitive interface areas in which machine and software could change their behavior based on the users' response [14, 15]. Affective computing is an emergent area of research which deals with the relationship between the emotions and computers [15]. It's a branch of computer science and it carries out the process of designing and developing technologies that recognize, express and understand the humans' emotions to better serve people's desires and to improve communication between people especially those with special needs [14][16][17]. The main area of affective computing as Picard said is the feelings that the machines might actually "have" [14]. Psychology, cognitive, physiology and computer science are the disciplines that are associated with affective computing [18]. Therefore, emotions and related human behaviors are indispensable backgrounds of affective computing [15].

Emotions are physiological changes that take place in the body. Understanding others' emotions is often conducted by knowing them, making a conversation and sharing experiences and feelings with each other [15]. Obviously, individuals' awareness of others' emotions is seen to be more on their relatives and friends than strangers. In 2003, Picard linked emotions with weather metaphors in her study; it's hard to measure and predict the emotions as weather conditions. For instance, weather forecasting helps people from getting wet in a rainy day because they can know beforehand that they have to hold their umbrellas [14]. Therefore, knowing users' emotions can help in predicting certain situations. With the different emotions that humans encountered every day, the individuals' ability to report those momentum feelings differs in the extent of the personal characteristics and understanding of how they describe what they are feeling and whether or not it's accurate. Collecting these emotions have been studied for more than hundred years [16]. Skin conductivity, electrodermal activity (EDA), body temperature and heart rate (HR) are all considered as physiological measures that can be used to detect the emotional responses of individuals.

The proliferation of technologies that can measure, communicate and transform emotions such as computers, sensors and smartphones have been addressed in several research studies [e.g. 22, 27, and 28]. A recent research study has been made using Q Sensor on people with severe mental disability who couldn't express their feelings in a verbal language to understand their emotional reactivity variations [23]. Besides, [26] proved that people can express what they feel with some distinct levels of variations in the expression. In another study, a Remote Millimeter Wave I–Q Sensor was used in a study to analyze the real-time heart rate beats which revealed that the tool and the method were both successful in gathering and detecting the heart beats and beat-to-beat heart rate even in places with different levels of noise. Thus, it can assist other applications in their detection of many kinds of heart diseases [24]. In addition, with technology changing rapidly, nowadays it can be possible to assess the persons' HR by using the smartphones' cameras. What is more, Lakens, D has shown that the method was successful in determining the anger and happiness feelings which the individuals express during experiments [22]. Table 1 illustrates additional technologies used in other experimental studies to detect different emotions.

Table 1 Methodologies used in several studies to detect emotions

| Study | Technology | What it measure? | Stimulus | Emotion recognition | Result |
|---|---|---|---|---|---|
| Gathering data which is represented by the bright that emit from the device of each participant to measure the person's emotions [19]. | Galvactivator | Skin conductivity | Listening to a session while wearing the Galvactivator gloves. | Physiological arousal. | The light is highly bright at the beginning of the presentation and in interactive sessions. Whereas, the light brightness is low at the end of the session. |
| An approach has been proposed for recognizing emotions based on physiological signals [20]. | ProComp | Electromyogram signals (EMG), Respiratory Volume (RV), skin temperature (SKT), skin conductance (SKC), blood volume pulse (BVP) and heart rate (HR). | Presentation of IAPS Pictures (International Affecting Picture System). | Joy, sadness, fear, disgust, neutrality an amusement. | Recognition rate of 85% for different emotional states. |
| Examining computational emotions recognition using multi-model bio-potential signals [21]. | Two sensors: pulse sensor clip and skin conductance electrodes. | Pulse and skin conductance. | Audio content. For negative emotions: engine sound of a motorcycle, alarm sound of an alarm clock, noise in a factory and cutting/breaking sound. For positive emotions: two tunes were selected and the tunes were based on subject's choice. | *Positive emotions:* relax and pleasure. *Negative emotions:* stressful and unpleasure. *Normal emotion.* | The recognition rate of 41.2% has been attained for the three emotions' categories. |
| Investigating electrodermal activity (EDA) and temperature [17]. | Q Sensor | Skin conductivity, temperature and motion. | Four different videos one for each selected emotion. | Fear, Happiness, sadness and disgust. | Many emotional arousals for each of the three subjects. |

# 3. METHODS

## 3.1 Participants

This research involved gathering data from 30 female adults with ages ranging between 19 and 27 years old (Mean = 21.4 years and Standard deviation (SD) = 2.04 years). The detailed ages' groups for the participants are depicted in table 2.

Table 2 Participants Demographics

| Age | # of participants |
|---|---|
| 19 | 8 |
| 20 | 2 |
| 21 | 8 |
| 22 | 1 |
| 23 | 7 |
| 24 | 3 |
| 27 | 1 |
| **Total** | 30 |
| **Mean** | 21.4 |
| **SD** | 2.0443 |

Participants were faculty and students in the university environment and they were informed about the experiment through announcements posted on campus. Also online announcements were posted to reach out to more audience. The study was conducted in three days and participants were allocated 30 minutes session for each experiment.

## 3.2 Apparatus

Several software and hardware devices were used in this study. LED and TV. Firstly, the Q Sensor1 (figure 1) which is a non-obtrusive bracelet worn on the wrist. It is a biosensor that measures the emotional arousal wirelessly via skin conductance that is a form of Electrodermal Activity (EDA). Moreover, EDA is an electrical changes measured at the surface of the skin that arise when the skin receives innervating signals from the brain [2] that increased or decreased during situations such as excitement, attention, anxiety, boredom or relaxation. Q Sensor measures temperature and activity of persons' bodies [6]. Secondly, a controller-free and full body gaming device Xbox Kinect (figure 2). Its sensors capture persons' gestures and respond to them.

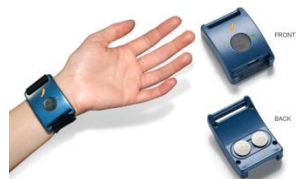

Figure 1 Q Sensor Device [4]

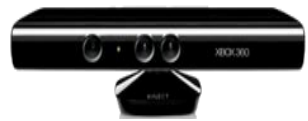

Figure 2 Kinect Device [5]

---

[1] "The Q Sensor has been tested to comply with IEC61010–1:2001, EN61010-1:2001 safety standards for laboratory equipment, and with relevant product safety standards for North America and Europe." [3]

Lastly, QLive was used for the purpose of viewing, tracking and recording a live stream of data from Q Sensor which appears as graphs and to ensure that there is no connection errors.

### 3.3  Stimuli

A chosen game named "Adventure Game" was used as stimuli for this experiment [25]. This game consists of two levels with approximate duration of five minutes.

### 3.4  Procedure

The experiment was conducted in a room within a sufficient space for capturing the participant's movement (figure 3).

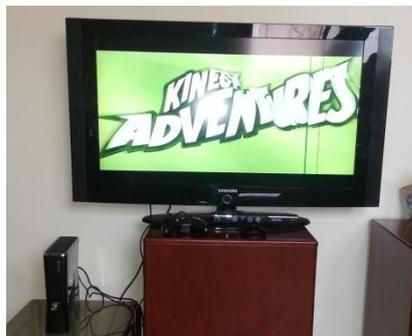

Figure 3 Experiment Room

In this experiment, each participant was requested to perform a task consisting of two phases. In the first phase, the participant was asked to warm-up by walking up and down the stairs for three minutes to ensure skin conductance responses were detected by the QSensor device. After that, she was requested to relax for six minutes. In the second phase, the participant was asked to wear the Q Sensor bracelet on the right hand's wrist and stand in front of the Xbox Kinect to play a game. The "Adventure game" was presented to the user as shown in (figure 4).

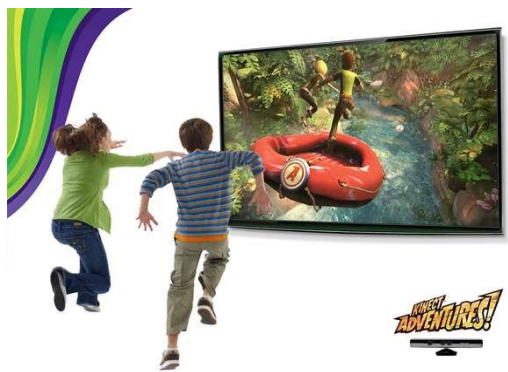

Figure 4 Adventure game [30]

Lastly, a game experience survey was presented to participants be filled. During the second phase, there were two observers whom carry the role of specifying the level of excitement of the player each minute through a pre-designed scale and a dedicated facilitator was working on the QLive software that was activated during sessions. As an incentive, a fifty Saudi riyal coupon was presented to each participant as well as a certificate of participation.

### 3.5  Data Analysis

#### 3.5.1.  Q sensor's Data Analysis

The Q Sensor generates its data in graphs as shown in figure 5. These graphs were taken and divided into four/five segments based on the participant's playing time. Therefore, each of these segments represents the participant's interactivity with the game for a duration of one minute. Consequently, the highest peak value for each part was specified and the average of these peaks was calculated accordingly as depicted in table 3. To determine the participant's level of excitement, the means of EDA in the specified segment was mapped to either excited, moderate or not excited. By observing the obtained averages, the range was between 0.11μs and 7.776μs and on average the mean of EDA was 3.7μs. According to the range between the minimum and the maximum values, this area was equally divided into three sub ranges which represent the three levels of excitements during the game. As a result, the average between 5.34μs and 7.776μs was mapped to be excited, the range between 2.67μs and 5.33μs indicates that the participant was moderately excited, and the not excited level is located in the area between 0.11μs and 2.665μs (table 3).

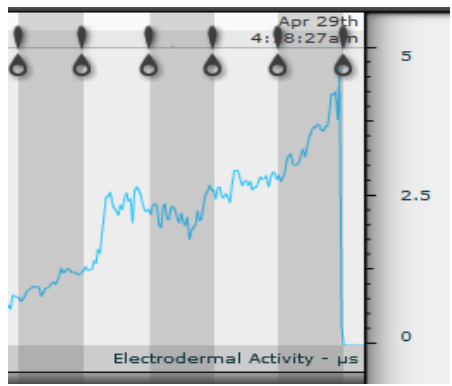

Figure 5 Q sensor graph for one participant

Table 3 Sample of 5 Participants

| ID | 1st minute (μs) | 2nd minute (μs) | 3rd minute (μs) | 4th minute (μs) | 5th minute (μs) | Average (μs) | Q Sensor Result |
|---|---|---|---|---|---|---|---|
| 1 | 3.31 | 3.64 | 4.06 | 4.39 | 4.56 | 3.992 | Moderate |
| 2 | 4.05 | 4.14 | 4.36 | 5.19 | 5.93 | 4.734 | Moderate |
| 3 | 1.46 | 1.34 | 1.7 | 1.78 | 0 | 1.57 | Not Excited |
| 4 | 0.42 | 1.5 | 1.45 | 1.41 | 1.52 | 1.26 | Not Excited |
| 5 | 3.19 | 4.39 | 5.07 | 6.29 | 6.94 | 5.176 | Excited |

After applying this methodology to all the 30 participants' data, one outlier with an EDA average of 15.83μs shown in table 4 was found and hence it was removed from our sample.

Table 4 Outlier data

|  | 1st min | 2nd min | 3rd min | 4th min | 5th min | Average |
|---|---|---|---|---|---|---|
| EDA | 10.77 μs | 15.46 μs | 16.53 μs | 17.29 μs | 19.11 μs | 15.83 μs |

### 3.5.2. Observation

Direct observation was used as it has the highest degree of 'ecological' validity [29]. in each session two observers simultaneously observed the 30 participants and specified the level of excitement for each minute while the participants were playing. Table 5 shows a sample of the data recorded by one observer. Moreover, the observers have agreed on the criteria for determining the excitement level beforehand based on the facial expressions such as smiling, eyes focus and body movements as feet bouncing. According to these criteria, big smiling and laughing were considered as marks to indicate that the participant was excited, whereas slight smiling and slow body movements show that the excitement level was moderate, and the boredom and laziness appearance in the participant's facial expressions or body movements is mapped to the not excited level. The data from the two observers for each session has been compared.

Table 5 Observation sample for one participant

| Level of excitement | 1st | 2nd | 3rd | 4th | 5th |
|---|---|---|---|---|---|
| Excited |  | √ | √ | √ |  |
| Moderate | √ |  |  |  | √ |
| Not excited |  |  |  |  |  |

After the indication of the excitement level for each minute, the level of excitement for the whole activity was determined by calculating the median. To estimate the median, number one was assigned to "Not excited", number two to "Moderate" and number three to "Excited". Therefore, four/five assigned numbers were sorted in ascending order and by taking the average of the middle located numbers if there are four assigned numbers or the third number if there are five assigned numbers to specify the value of the median. Then, this result was matched to the appropriate excitement level to get the overall scale of enjoyment (table 6).

Table 6 Sample of observers' results for 5 participants

| ID | Observer# 1 | Observer# 2 |
|---|---|---|
| 1 | Moderate | Moderate |
| 2 | Not Excited | Moderate |
| 3 | Excited | Excited |
| 4 | Moderate | Moderate |
| 5 | Moderate | Moderate |

In addition, in order to avoid the participants' nervousness of observers' presence, they have spent their six minutes' relaxation period. The goal of this period was to let the participants become less conscious of the observers'[29]. The participants were informed at the beginning of the session about the purpose of the study, the reason of documenting their activities and the investigation which took a place while they were playing.

### 3.5.3. Survey

At the end of each session, a game experience survey was given to the participant to evaluate her excitement level. This survey contains seven various questions which gather information about their demographics, their excitement level and their experience in the game. The latter two points will be illustrated below.

The level of excitement of the participant was determined from what she has reported about herself in the survey to use it as a subjective measure (table 7). The answers were taken by us in order to map them to either extremely excited; moderately excited; or slightly excited. Table 8 shows how this mapping was performed. A sample of five participants' excitement level is depicted in table 9.

Table 7 Question# 6 from the game experience survey

| The survey question used to measure the excitement level of the participant (English) | The survey question used to measure the excitement level of the participant (Arabic) |
|---|---|
| 6- Did you feel excited while playing the game?<br>☑ Yes.<br>o No.<br><br>1) If yes, for what level do you rate your excitement?<br>    o Extremely.<br>    ☑ Moderately.<br>    o Slightly. | 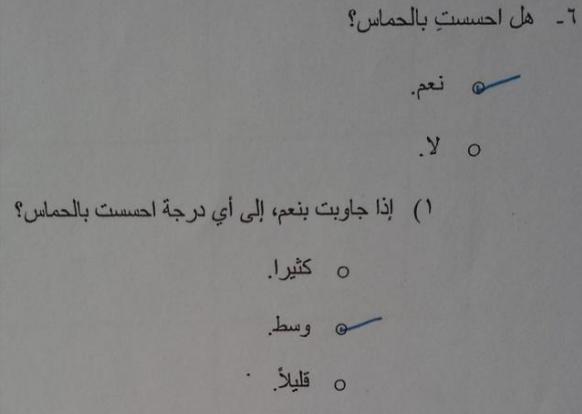 |

Table 8 Self-reported Data Mapping

| Question# 6 answer | Question# 6.1 answer | Result |
|---|---|---|
| Yes | Extremely | Excited |
| Yes | Moderately | Moderate |
| Yes | Slightly | Not excited |
| No | - | Not excited |

Table 9 Sample of Survey results for 5 Participants

| ID | Survey Results |
|---|---|
| 1 | Moderate |
| 2 | Excited |
| 3 | Excited |
| 4 | Moderate |
| 5 | Excited |

### 3.5.4. Demographic Survey Related to Gaming

As for the participants' previous experience with the game, nineteen participants didn't play the game before, while ten of them did. Those ten participants were divided into three categories. Six of them have played the game once before, two of them have played it two/three times and the remaining two participants have played it for five times or more (Mean = 0.655 µs and (SD) = 3.566 µs).

Table 10 Question#7 from the game experience survey

| The survey question used to indicate the participant's previous experience (English) | The survey question used to indicate the participant's previous experience (Arabic) |
|---|---|
| 7- Did you play the game before?<br>    o Yes<br>    √ No<br><br>    1) If yes, for how many times?<br>        o One time.<br>        o From two to three times.<br>        o Five or more. | ٧- هل لعبتها مسبقاً؟<br>    o نعم.<br>    o لا.<br><br>    ١) إذا جاوبت بنعم، كم مره لعبتها؟<br>        o مره واحده.<br>        o مرتين الى ثلاث مرات.<br>        o ٥ مرات فأكثر. |

## 4. Results

Different kinds of data to determine the excitement level have been gathered in the previous sections which came from many sources. These are the EDA-based physiological sensor generated graphs, observers' filled forms and the game experience surveys. In order to analyze these collected data, the data has been studied from two different angles. Firstly, examining the relationship between the previous playing times collected from the game experience surveys and the excitement level that was specified by either the self-reported data or the EDA-based physiological sensor data. Secondly, statistical analysis that were performed by using IBM SPSS predictive analytics software.

### 4.2. Survey Results

According to the surveys' results, figure 6 shows the relationship between the number of participants' previous playing times and their excitement level based on their self-reported data. This was mainly done to know how did they feel during the game. This chart demonstrates that despite the fact that there are nineteen participants who didn't play the game before, 47.36% of them have reported that they were excited whereas only 5.26% indicated that they weren't and the remaining participants have reported that they were moderately excited. Moreover, there are six participants who have played the game once before, 83.33% of them stated that they were excited while 16.76% of them were moderately excited. Furthermore, the last two categories have got the same number of participants which is two per each. One of the two participants who have played the game two to three times before reported to be extremely excited, while the other was moderately excited. For the participants who have played the game five or more times before, all of them were moderately excited.

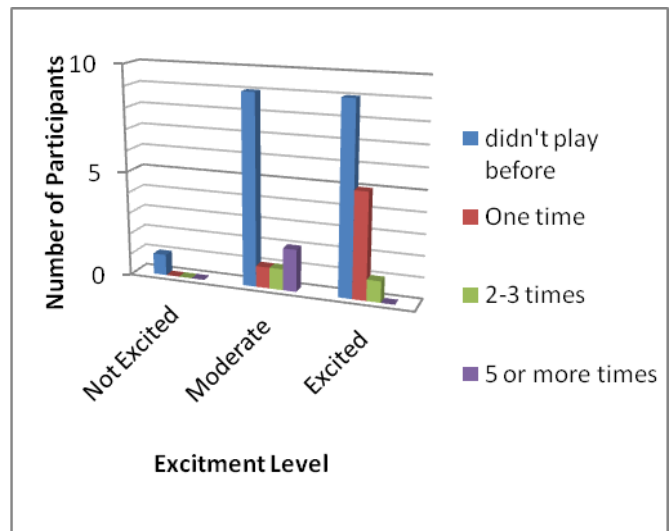
Figure 6 Game experience and excitement level (survey)

On the other hand, figure 7 illustrates the relationship between the number of playing time collected from the surveys and the excitement levels obtained from the EDA-based physiological sensor results. Based on the EDA-based physiological sensor observation, the chart below shows that 16.7% were excited, 20% were not, and 26.7% were located in the middle. All of the nineteen participants didn't play the game before. In addition, for the participants who have played the game one time before, EDA-based physiological sensor analysis indicated that the excitement level for two participants was moderate, the same number were not excited and two participant were extremely excited. What is more, all of the participants who have experienced the game two to three times before weren't excited and all of the subjects who have played the game five or more times were moderately excited.

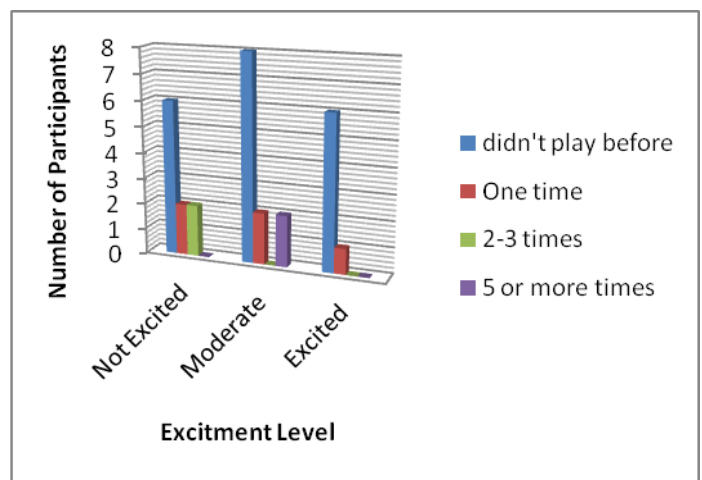
Figure 7 Game experience and excitement level (Q Sensor)

### 4.3. Tests Based Analysis

In order to examine if our findings as demonstrated in Table 11 can support our hypothesis to use the EDA-based physiological sensor device as usability testing tool, a manual calculations along with statistical tests have been used.

Table 11 Sample of observers' data, surveys' data and Q sensor's results for 5 Participants

| ID | Observer #1 Results | Observer #2 Results | Survey Results | Q Sensor Results |
|---|---|---|---|---|
| 1 | Moderate | Moderate | Moderate | Moderate |
| 2 | Not Excited | Moderate | Excited | Moderate |
| 3 | Excited | Excited | Excited | Not Excited |
| 4 | Moderate | Moderate | Moderate | Not Excited |
| 5 | Moderate | Moderate | Excited | Excited |

### 4.3.1. Comparative analysis of results:

In this section, the level of agreement between the outcomes of the observers, EDA sensor and the game experience survey results are presented. The following sections address the way in which each comparison was performed.

#### 4.3.1.1. External Observation

The percentage of agreement between the two observers' results is 79.31%. This result came from the fact that there were twenty-three matching results out of twenty-nine, and this percentage of agreement is relatively high.

#### 4.3.1.2. External Observation and the Survey Results

The percentage of agreement between the two observers' results and the surveys' results is 44.83%. This result came from the fact that there were thirteen matching results out of twenty-nine, and this percentage of agreement seems to be fair.

#### 4.3.1.3. External Observation and the EDA-Based Sensor Results

The percentage of agreement between the two observers' and the EDA-based sensor findings is 31.03%. This result came from the fact that there were nine matching results out of twenty-nine, and this finding is relatively low.

#### 4.3.1.4. The Survey and the EDA-Based Sensor Results

The percentage of agreement between the survey results and the EDA-based sensor findings is 34.48%. This result came from the fact that there were ten matching results out of twenty-nine. Similar to the External Observation and EDA-Based Sensor Results this finding is relatively low.

#### 4.3.1.5. External Observation, Survey and EDA-Based Sensor Results

The percentage of agreement all of the subjective and objective measures is 10.35%. This result came from the fact that there were three matching results out of twenty-nine, and this percentage of agreement seems to be very low.

### 4.3.2. Statistical Tests

Statistical tests have been conducted, these tests are Wilcoxon signed ranks test and Friedman test. In the tests, a P-value will be generated to check if there's a significant difference between the entities by comparing the P-value with the constant number 0.05. If the P-value is higher than or equal to 0.05 then there is no significant difference between the entities, otherwise there is a difference.

#### 4.3.2.1. External Observation

Since there are two observers acting as one parried group, Wilcoxon signed ranks test has been used and the measurement was ranked as excited, moderate and not excited. The Wilcoxon signed ranks test revealed that there is no statistical differences between the two observers (P-value = 0.102 which is greater than 0.05).

#### 4.3.2.2. External Observation and Survey Results

Friedman Test has been used because it is normally conducted when there is the same sample of participants and the measurement is taking place at three or more points in the same time, which are the two observers and the survey. The results obtained fro this test indicated that there is no significant difference was evident between the two observers and the survey result (P-value= 0.199, which is greater than 0.05).

#### 4.3.2.3. External Observation and the EDA-Based Sensor Results

There is a significant difference between the observers and the EDA-based sensor results when using Friedman test (P-value= 0.003, which is less than 0.05).

#### 4.3.2.4. Survey and the EDA-Based Sensor Results

Using Wilcoxon signed ranks test, the results clearly showed that there is a significant difference between the survey and the EDA-based sensor results (P-value= 0.007, which is less than 0.05).

#### 4.3.2.5. External Observation, Survey and the EDA-Based Sensor Results

High difference has been revealed between the observers, survey and the EDA-based sensor results when using Friedman test (P-value= 0.001, which is less than 0.05).

## 5. DISCUSSION

To find a relationship between the excitement level of the participants and their previous experience in playing the "Adventure Game"; the EDA-based sensor data and self-reported data were examined. As a result, those who didn't play the game before, showed 31.58% matching between the subjective and objective measure sources. This finding comes from the fact that there were six participants out of nineteen whose results in both sources indicated an equal category of being extremely excited, moderately excited. Moreover, there were two out of six matched participants from those who played the game once before showed a matching percentage of 33.33%. There wasn't any match between EDA-based sensor results and the surveys' results for those who have played the game two to three times before. In contrast, the result for the participants who have experienced the game before, matched perfectly between the two measures. Measuring the user experience of the participants who were familiar with the game, was effective when capturing it with the EDA-based sensor data. And this because it was in line with the observation and in line with the self-reported data. On the contrary, the users whom were unfamiliar with the game results varies.

Generally, without considering the external observation, the agreement percentage between Q Sensor's results and surveys results was 34.48%. According to the test based analysis, 79% was the agreement percentage between the two observers which drive us to believe that they almost have the same vision, and Wilcoxon signed ranks test proves this finding by revealing that there are no statistical differences between them. After that, the agreement percentage between the observers and the surveys' results was calculated and we found that they have agreed to the ratio of 44%, while the Friedman test indicated that there is no significant difference was evident between them. Consequently, we found that the observers were partially able to sense the participants' feelings and therefore the two subjective measures used in this research study agreed with each other to a large extent. Turning to consider the objective measure used in this study, which is the EDA-based sensor data and by comparing it with the observers' data, we found that the percentage of agreement between them was low in conjunction with the Friedman test as it showed that there was a significant difference between them. Since this percentage is relatively low, we have decided to consider the second subjective measure which was the self-reported data and the agreement percentage was also low. Furthermore, conducting Wilcoxon signed ranks test revealed that there was a significant difference between them. The findings suggest that the EDA-based sensor didn't effectively measure the level of excitement of the user when compared to self reported data and human observers rating.

From this experience it would be interesting to consider the overall agreement in a triangular approach similar to the approach presented in [13]. This has been adopted to take into consideration three measures which are the external observation, self-reported data from the collected surveys and the objective measure which is the EDA-based sensor data. The agreement percentage between external observation, self-reported data and EDA-based sensor data were very low as the Friedman test showed.

These findings partially supported our hypothesis in that the EDA-based sensor was able to add insights to the user experience specially with people who are familiar with the system. However, the hypothesis wasn't supported when the EDA-based sensor was compared to observers' data and self-reported data because of the discrepancy between human reported data and self reported data from one side and the EDA-based sensor from the other side.

Another reason could be the individual differences between people and how the EDA-based sensor detecting this arousal level. An individual's reactions in specific situation differ from the reactions of other individual under the same circumstances.

## 6. CONCLUSION

This paper examined the use of the EDA-based sensor as a usability testing tool by performing an experiment on female adults. This experiment involves subjective and objective observations on subjects while they are wearing EDA-based sensor and playing a game named "Adventure Game" on Xbox Kinect device. Two perspectives were considered when analyzing the gathered data from observers, EDA-based sensor and surveys. First, finding a relationship between the previous experiences on the game and the excitement level that were specified based on one of the two sources which are EDA-based sensor data or self-reported data. Second, The analysis has been conducted on the collected data while taking into consideration the different combinations of the subjective and objective measures used in this study. As a result, the findings partially supported the hypothesis.

## 7. FUTURE WORK

Would examine the finding on a larger data set and perhaps with different bio sensors to see what other tools might be effective in Usability Testing in the context of game design and development.

## 8. LIMITATION

1. Fifteen minutes were set between the sessions to solve the Bluetooth connection problems of the Q Sensor with the laptop.
2. Due to the Q Sensor's discontinuation, the analysis feature wasn't supported anymore and we were forced to analyze the data manually and using IBM SPSS predictive analytics software.